\documentclass[3p, singlecolumn, fleqn]{elsarticle}

\usepackage{graphics}
\usepackage{epsfig}
\usepackage{epstopdf}
\usepackage{multirow}
\usepackage{amssymb}
\usepackage{amsthm}
\usepackage{gensymb} 
\usepackage{float}
\usepackage{lineno}
\usepackage{subfigure}
\usepackage[version=3]{mhchem} 

\journal{International Journal of Hydrogen Energy}

\begin{document}
\begin{frontmatter}

\title{Strain engineering and photocatalytic application of single-layer ReS$_2$}

\author[label1]{Yan-Ling Li}
\ead{ylli@jsnu.edu.cn}
\author[label2]{Yunguo Li}
\ead{yunguo.li@ucl.ac.uk}

\author[label1]{Chunlin Tang}

\address[label1]{School of Physics and Electronic Engineering, Jiangsu Normal University, Xuzhou, People's Republic of China}
\address[label2]{Department of Earth Sciences, University College London, Gower Street, London WC1E 6BT, United Kingdom}


\begin{abstract}
We present a theoretical study on the electronic, dynamical, and photocatalytic properties of single-layer ReS$_2$ under uniaxial and shear strains. The single-layer ReS$_2$ shows strong anisotropic responses to straining. It remains dynamically stable for a wide range of $x$-axial strain, but becomes unstable for 2\% $y$-axial compressive strain. The single-layer ReS$_2$ is calculated to be an indirect bandgap semiconductor, and there is an indirect$-$direct bandgap transition under 1$-$5\% $x$-axial tensile straining. The single-layer ReS$_2$ is predicted incapable of catalyzing the water oxidation reaction. However, 1$-$5\% $y$-axial tensile strain can enable the single-layer ReS$_2$ for overall photocatalytic water splitting. Besides, the single-layer ReS$_2$ can also catalyze the overall water splitting and be most efficient under acidic water solutions with pH=3.8.

\end{abstract}


\end{frontmatter}


\section{Introduction\label{I}}
Hydrogen, sourced from photocatalytic water splitting, is one of the most promising methods to provide energy for the sustainable development of our economy and society.
Qualified photocatalysts have the bandgap over 1.23 eV. The conduction band minimum (CBM) should be more negative than the H$^+$/H$_2$ water reduction potential, while the valence band maximum (VBM) should be more positive than the OH$^-$/O$_2$ water oxidation potential \cite{rsc12lawerenz}.

Nanostructured materials have shown their advantages in photocatalytic applications \cite{ANIE:ANIE201306918,ANIE:ANIE201410172,doi:10.1021/ja506261t}. Bandgap can be opened up due to the quantum confinement effect in nano-sized materials, and the increased surface-to-volume ratio provides more chemical-active sites \cite{ANIE:ANIE201306918,C4CS00236A}. These can effectively improve the photocatalytic efficiency. Besides, nanostructured materials coupled with traditional or other nanostructured materials, can further enhance the catalytic properties via the mechanisms of Z-scheme, plasmonic resonance effect and so on \cite{MRS:7966834,jacs08awazu}. Among the nanostructured materials, two-dimensional (2D) materials is more efficient because of the best surface-to-volume ratio and abundant chemical-active sites \cite{C4CS00236A}. 

The single-layer MX$_2$ (M=Mo, W, Nb; X=S, Se, Te) is a series of typical 2D materials that have promising applications in photocatalytic water splitting. The bulk MX$_2$ is usually indirect bandgap semiconductor \cite{Li2014206}. There is an indirect$-$direct bandgap transition and improved light absorption when thinning the bulk MX$_2$ into single layers \cite{alvarez2015single,catal13li}. 
Some of the single layers are semiconductors with the bandgap between 1.1 eV and 2.1 eV \cite{Li2014206}. MoS$_2$, MoSe$_2$, and WS$_2$ have been shown to produce high yields of H$_2$ \cite{ANIE:ANIE201306918,aplmater14gupta,doi:10.1021/ja506261t}. Three types of structure have been observed among the MX$_2$ single layers: 1H-MX$_2$ (space group \emph{P$\bar{6}$m$2$}), 1T-MX$_2$ ( space group \emph{P$\bar{3}$m$1$}), and distorted 1T-MX$_2$ (space group \emph{P}-1). The metal atoms are trigonal prismatically coordinated in the 1H-MX$_2$ structure, and octahedrally coordinated in the 1T-MX$_2$ structure. The distorted 1T-MX$_2$ is a triclinic structure with distorted X octahedron. It was recently found that the single-layer ReS$_2$ keeps the distorted 1T-MX$_2$ structure as in the bulk ReS$_2$ \cite{tongay14natcomm}. It was proved that the ReS$_2$ layers are electronically and dynamically decoupled in the bulk with very weak van der Waals forces. The bulk and single layer possess nearly identical band structure and Raman active modes \cite{tongay14natcomm,PhysRevB.92.054110}. This unique character makes the bulk ReS$_2$ a pseudo single-layer material. This also creates great advantage for ReS$_2$ if it can be engineered for photocatalytic applications. The weak coupling between single layers in ReS$_2$ allows for easy preparation of nanosheets that can provide abundant surfaces and can be easily cocatalyzed with other materials.  

However, the potential photocatalytic application of the two-dimensional ReS$_2$ has not been comprehensively investigated. In this paper, we explore the photocatalytic application of the single-layer ReS$_2$ by using density functional theory (DFT) based calculations. First, we present the electronic and optical properties of the pristine single-layer ReS$_2$; Then we investigate the variations of the dynamic, electronic, and photocatalytic properties of the strained single-layer ReS$_2$. At last, we discuss the potential application of the single-layer ReS$_2$ as an photocatalyst .

\section{Methodology\label{II}} 

The present calculations are based on DFT and use a plane-wave basis set, as implemented in the Vienna Ab initio Simulation Package (VASP) \cite{kresse1999ultrasoft, blochl1994projector}. The interaction between the ions and valence electrons is described by the projector augmented wave (PAW) method \cite{kresse1996efficiency, kresse1996efficient, kresse1993ab}.  Most of the calculations, including structural relaxations and phonon spectra calculations, are done on the level of generalized gradient approximation (GGA), employing the exchange-correlation functional by Perdew, Burke and Ernzerhof (PBE) \cite{perdew1996generalized}. In general, semi-local GGA functionals are known to underestimate the electronic bandgap of semiconductors and insulators, while providing good structural accuracy \cite{PhysRevB.77.165107,catal13li,jpcc14sa}. The S-3$s^2$3$p^4$ and Re-5$d^5$6$s^2$ are treated as valence electrons. The convergence criteria is 1$\times$10$^{-6}$ eV/atom for the total energy and 1$\times$10$^{-5}$ eV/\AA for the force on atoms. We performed tests to find suitable cut-off energy and $\bf{k}$-mesh. A plane-wave cut-off energy of 500 eV and a $\bf{k}$-mesh of $16\times$16$\times$1 were enough to reach convergence and used in the calculations.  

The phonon dispersions were calculated by means of PHONOPY code \cite{PhysRevB.78.134106}, which is an implementation of post-process phonon analyzer, from the Hessian matrix calculated using density functional perturbation theory (DFPT) and PBE functional implemented in VASP. We used a 2$\times$2$\times$1 supercell to calculate the eigenvalues of the Hessian matrix. The phonon-related thermal properties of these compounds were then derived from the calculated phonon spectra. 

\begin{figure}[!h]
\centering
\includegraphics[width=0.8\textwidth]{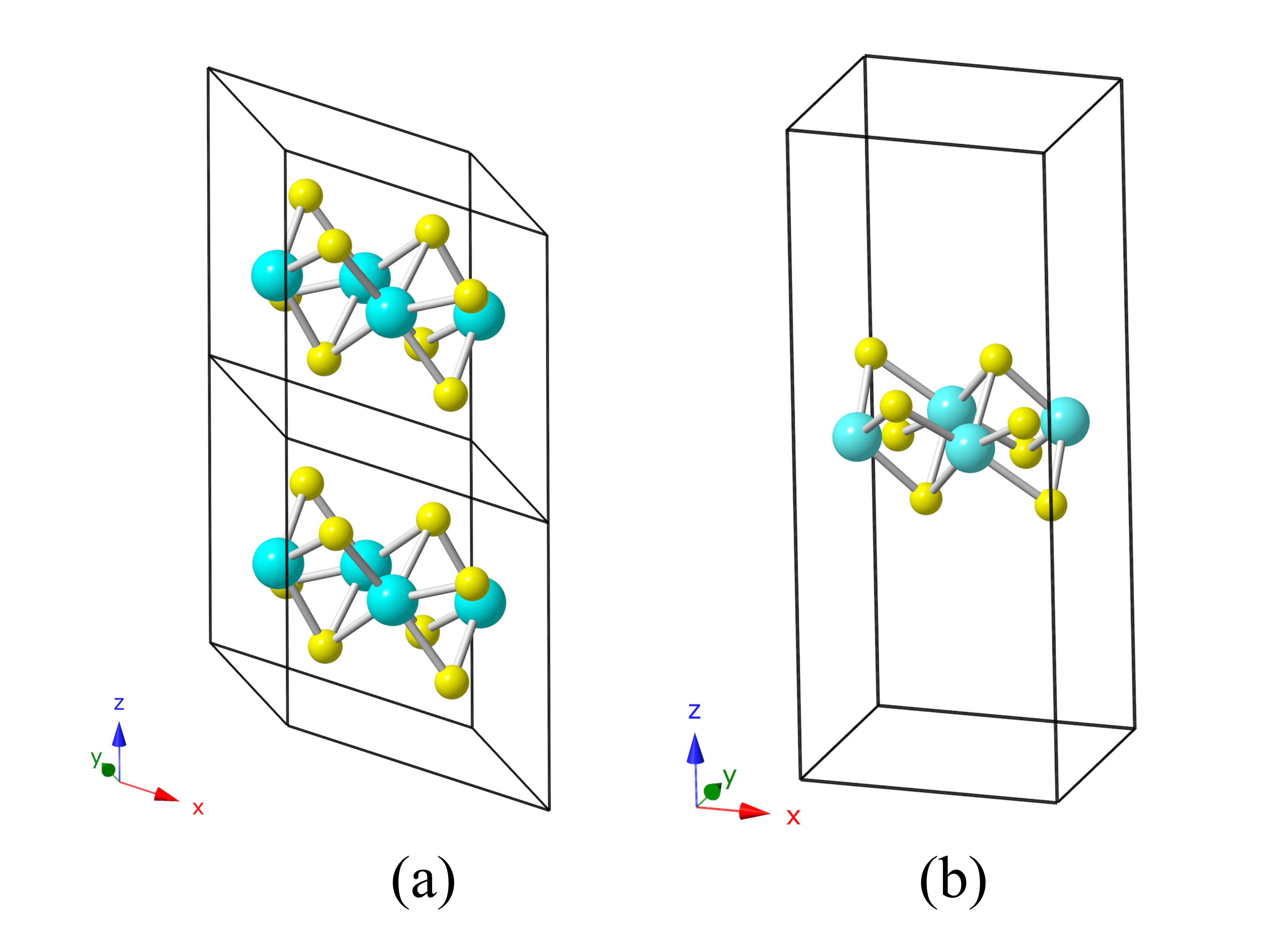}
\caption{The crystalline structures of (a) bulk and (b) single-layer ReS$_2$. The single layer is illustrated in an orthorhombic unitcell. Cyan and yellow balls indicate Re and S atoms, respectively.\label{fig:str}}
\end{figure}

The bulk ReS$_2$ corresponds to a triclinic structure, which can be considered as a transformed structure from the hexagonal lattice via Peierls distortion. The single layers in ReS$_2$ keeps the distorted 1T-MX$_2$ structure. To facilitate the simulation, we adopted an orthorhombic unitcell for the single layer. Figure \ref{fig:str} shows the structures of the bulk and single-layer ReS$_2$.
\begin{figure}[!h]
 \centering
\includegraphics[width=0.6\textwidth]{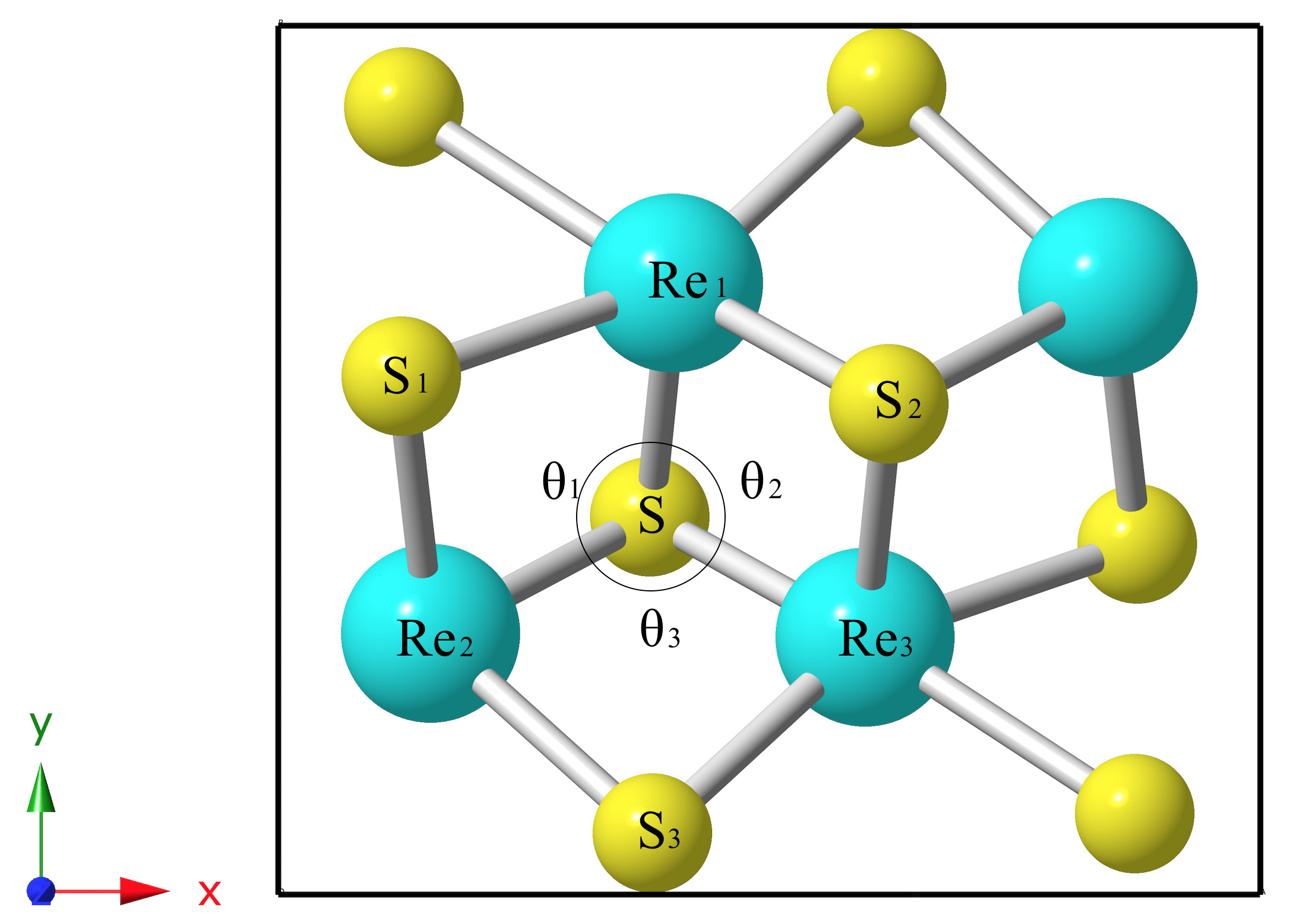}
\caption{The top view of the single-layer ReS$_2$ unitcell \label{fig:label}}
\end{figure}

The $x$-axial strain ($\varepsilon(x)$), $y$-axial strain ($\varepsilon(y)$), and shear strain ($\varepsilon(shear)$) were applied upon the relaxed ground-state structure, respectively. The transform matrix is

$\begin{pmatrix} 1+\varepsilon(x) & \varepsilon(shear)/2 & 0 \\  \varepsilon(shear)/2 & 1+\varepsilon(y) & 0 \\ 0 &0 & 1 \end{pmatrix}$

The internal atomic coordinates were fully relaxed for the strained structures, but he lattice vectors were not allowed to relax. This mostly resembles the case that single layers are located on substrates, where the lattice vectors of the single layer are locked respectively.

To better describe the changes of the structure, we labeled the atoms and the angles in the single-layer ReS$_2$ as depicted in Figure \ref{fig:label}.

\begin{figure}[!h]
\centering
\includegraphics[width=0.75\textwidth]{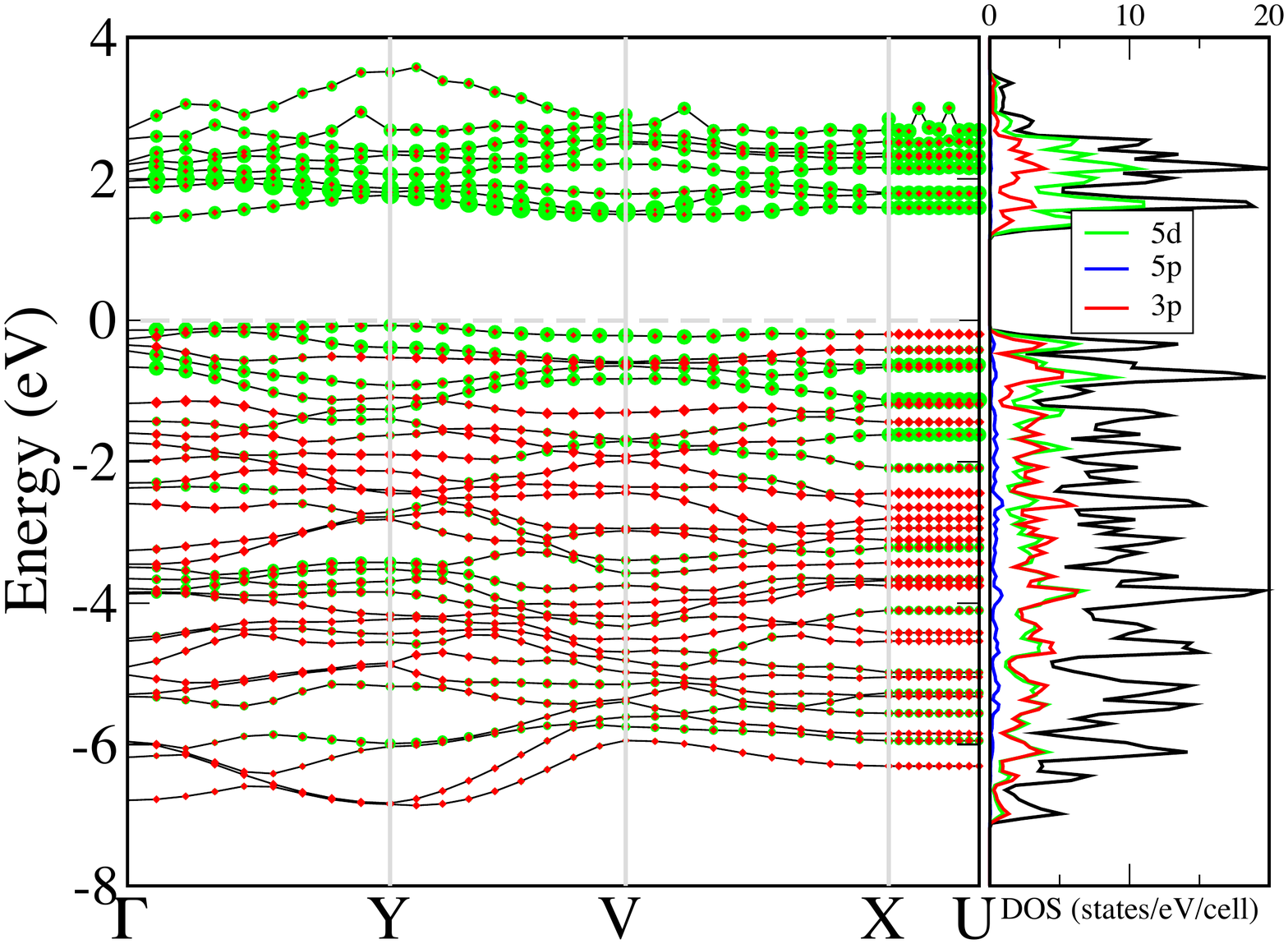}
\caption{The band structure and projected electronic density of states (DOS) of the single-layer ReS$_2$. The corresponding DOS are projected from the Kohn-Sham wave functions onto atomic Bader volumes and calculated within these volumes. \label{fig:band}}
\end{figure}
\begin{figure}[!h]
\centering
\includegraphics[width=0.7\textwidth]{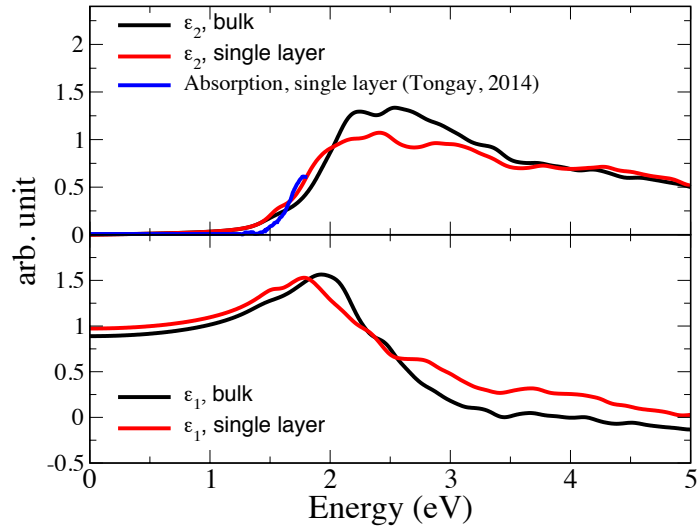}
\caption{Optical properties of the single-layer ReS$_2$. The upper panel and lower panel show the imaginary part $\varepsilon_2$ and real part $\varepsilon_1$ of the dynamic dielectric function, respectively. The experimental absorption spectrum is from Ref.\citenum{tongay14natcomm}. \label{fig:optic}}
\end{figure}

\section{Results and Discussion\label{III}}
\subsection{Ground-state single-layer ReS$_2$}
\begin{table*}[t]
 \centering
 \caption{Calculated lattice parameters of the single-layer ReS$_2$ in comparison with experimental data of bulk. \label{tbl:lattice}}
\begin{tabular} {c c c c c c c c}
\hline
  & a(\AA)  &b(\AA) & c(\AA) &$\alpha$($\degree$) &$\beta$($\degree$) &$\gamma$($\degree$) &Vol(\AA$^3$) \\
Bulk\cite{lamfers1996crystal} &6.352 &6.446 &12.779 &91.51 &105.17 &118.97 &434.49 \\
Bulk\cite{murray1994structure} &6.417 &6.510 &6.461 &121.10 &88.38 &106.47 &219.32 \\
Single layer &6.390 &5.726 &16.657 &90 &90 &90 &609.5 \\
\hline
 \end{tabular}
 \end{table*}
\begin{figure}[!h]
\centering
\includegraphics[width=0.6\textwidth]{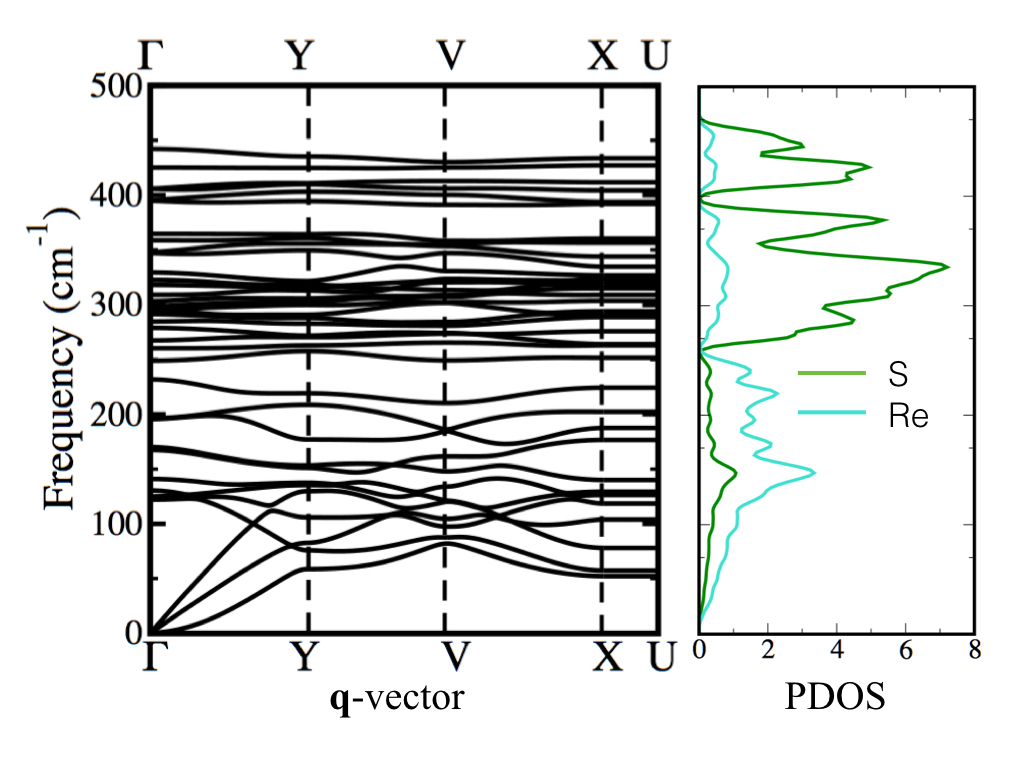}
\caption{The phonon spectra and partial density of states (PDOS) of single-layer ReS$_2$. \label{fig:ph}}
\end{figure}

Table \ref{tbl:lattice} shows the calculated lattice parameters of the single-layer ReS$_2$ in comparison with the bulk. The calculated lateral area of the single-layer ReS$_2$ is very close to that of bulk. Besides, the calculated Re--S bond length (2.33$\sim$2.44\AA) matches with that of experimental data and that of LDA (local density approximation) data \cite{PhysRevB.92.054110}. 

Besides, the calculated bandgap of the single-layer ReS$_2$ is 1.43 eV. This value is close to the experimental value (1.55 eV) and agrees with other calculations (1.43 eV) \cite{tongay14natcomm}. It suggests the good choice of our method. The bandgap calculated using HSE06 is 2.00 eV, which is highly overestimated. Such a phenomenon is usually seen in 2D semiconductors \cite{catal13li}. Hybrid functionals usually overestimate the bandgap of single layers, while the semi-local functionals can provide closer values to the experimental data. This can be attributed to the weak non-local correlations in 2D semiconductors. Besides, there is no sharp cutoff of the charge density on the surface and thus can be accurately described by the generalized gradient approximation. We also calculated the band structures of the bulk, four-layer, and two-layer ReS$_2$. They share much the same band structure. The calculated bandgap of the bulk ReS$_2$ is 1.36 eV, which is also very close to that of the single layer.

Our calculations show that both the single-layer and bulk ReS$_2$ are indirect bandgap semiconductors. The single-layer and bulk ReS$_2$ share a very similar band structure \cite{tongay14natcomm}. Figure \ref{fig:band} shows the band structure of the single-layer ReS$_2$. It was claimed that the single-layer and bulk ReS$_2$ are direct bandgap semiconductors, however, some recent high-quality experiments proved that the bulk ReS$_2$ is an indirect bandgap semiconductor \cite{aslan2015linearly,2d-res2-2016}. It is reasonable to expect an indirect bandgap character for the single-layer ReS$_2$, which is shown by our calculations and needs to be confirmed by experiments. 

As shown in Figure \ref{fig:band}, the single-layer ReS$_2$ exhibits very flat bands due to the low symmetry. There is a strong hybridization between the Re-$d$ states and S-$p$ states. This is similar to other transition metal dichalcogenides. The VBM, or the highest occupied crystal orbital (HOCO) is from the filled antibonding states of ($d_{xy}$, $d_{yz}$, $d_{zx}$) + ($p_x, p_y$) hybridization. The CBM, or the lowest unoccupied crystal orbital (LUCO) is from the empty antibonding of ($d_{z^2}$, $d_{x^2-y^2}$) + ($p_x, p_y$) hybridization. Our calculation shows that VBM appears at the $\bf{Y}$ point other than the $\bf{\Gamma}$ point \cite{tongay14natcomm,liu2016highly}. CBM appears at the $\bf{\Gamma}$ point in calculation, which is same to other calculations \cite{tongay14natcomm,liu2016highly}.

The calculated optical bandgaps of the bulk and single-layer ReS$_2$ are in agreement with their electronic bandgaps, as can be seen in Figure \ref{fig:optic}. The bulk and single-layer ReS$_2$ share very similar optical properties as expected. Their imaginary parts of the dynamic dielectric function almost coincide with each other, showing great similarity in the light absorption character. The calculated single-layer spectrum also matches the experimental spectrum of single layer \cite{tongay14natcomm}. It also can be seen that the absorption is more effective for photon energy ranging from 1.6 to 3.1 eV, which falls in the visible-light energy range. 

The single-layer ReS$_2$ is dynamically stable as there is no imaginary frequency in phonon spectrum shown in Figure \ref{fig:ph}. There are four formula units of ReS$_2$ in the single-layer unitcell as seen in Figure \ref{fig:str}. Therefore, it has 36 phonon branches which includes three acoustic branches and 33 optical branches. The phonon PDOS (partial density of states) suggests that the low-frequency region is mostly populated by Re. The high-frequency region, namely the optical branches is mainly dominated by S. This is attributed to the light weight and low coordination number of S. The three acoustic branches are not degenerate, suggesting anisotropic nature of the structure.

\subsection{Stability under strain}
\begin{figure}[!h]
\centering
\includegraphics[width=1\textwidth]{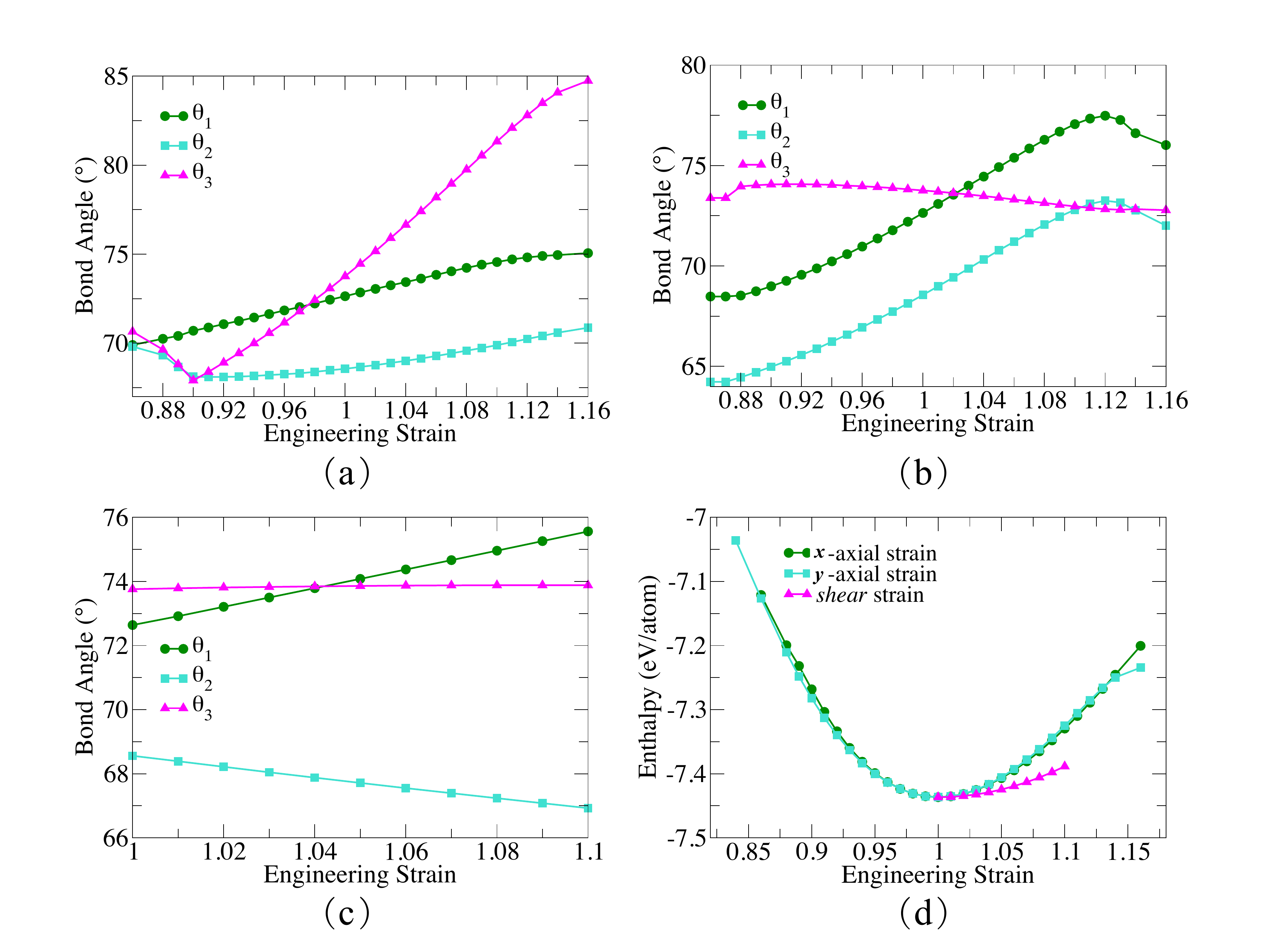}
\caption{The variations of bond angles as a function of (a) the $x$-axial strain, (b) $y$-axial strain, and (c) shear strain. The change of enthalpy is plotted in (d). \label{fig:angle}}
\end{figure}
\begin{figure}[!h]
\centering
\includegraphics[width=0.8\textwidth]{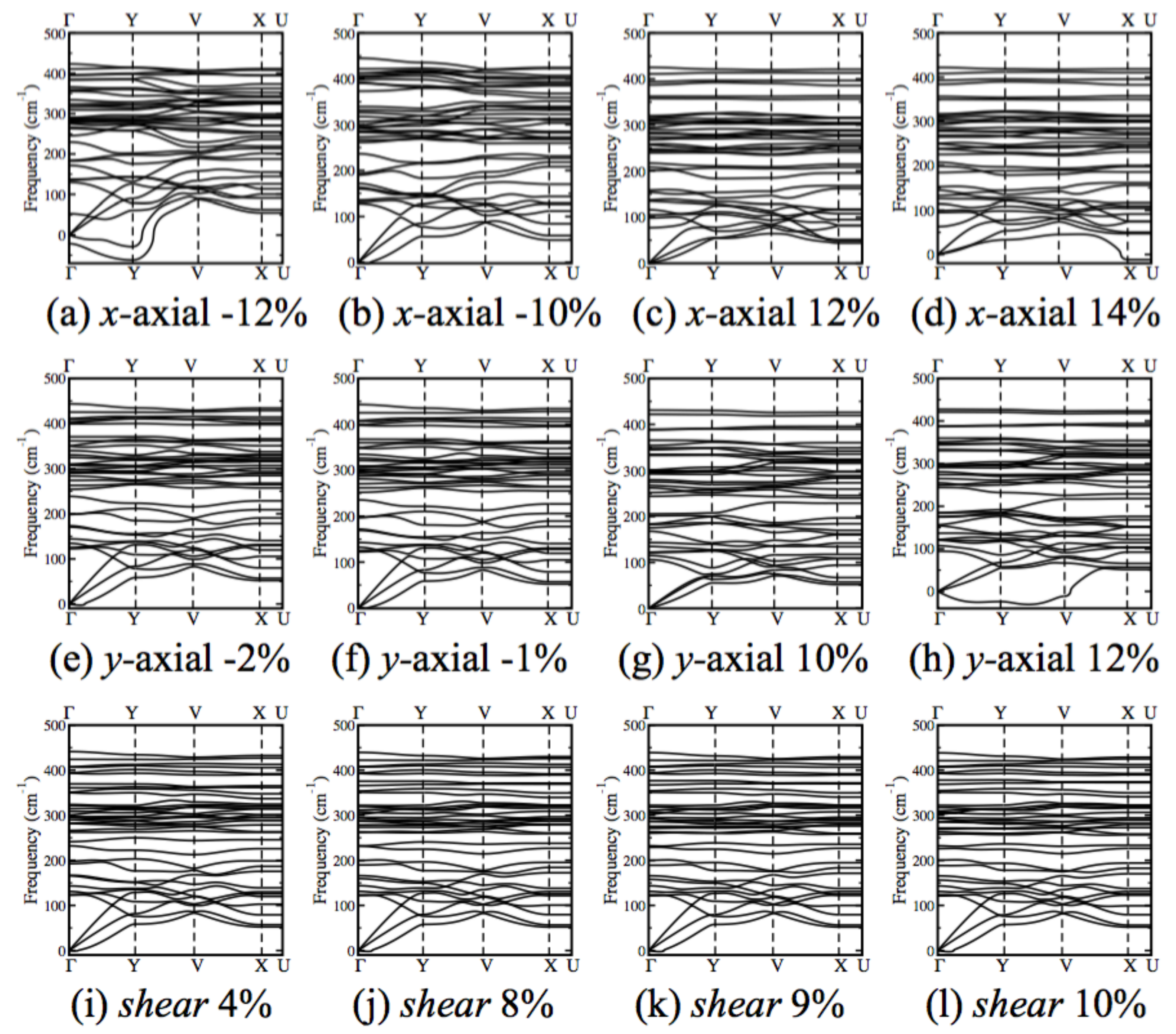}
\caption{Phonon dispersion curves of single-layer ReS$_2$ under strain. \label{fig:phonon}}
\end{figure}
In this section, we try to explore the elastic limit of the single-layer ReS$_2$. The uniaxial strains in $x$-axial and $y$-axial directions and shear strain were applied to the relaxed ground-state structure. The internal atomic coordinates were fully relaxed after applying the engineering strain.

Figure \ref{fig:angle} shows the changes of bond angles under uniaxial strains along $x$-axis and $y$-axis. The bisector of angle $\theta_3$ is almost parallel to the $y$-axis, so the change of $\theta_3$ under the $y$-axial strain is opposite to that under the $x$-axial strain. The bisectors of angle $\theta_1$ and $\theta_2$ lie between the $x$-axis and $y$-axis, so the responses of $\theta_1$ and $\theta_2$ under uniaxial strains exhibit similar trend. Similarly, according to the geometry, $\theta_1$ and $\theta_2$ will show opposite responses to the shear strain, while  $\theta_3$ will be hardly affected, as shown in \ref{fig:angle}.

There is no abrupt change of enthalpy under the examined $x$-axial and shear strain, but the enthalpy shows an increased drop for the $y$-axial tensile strain over 10\% and a non-smooth change over 13\%. The angles also show abnormal changes over 10\% $y$-axial tensile strain or over -12\% $y$-axial compressive strain. Under the $x$-axial strain, the angles change abnormally overn 12\% tensile strain and over -10\% compressive strain. There is no abrupt changes of angle under the shear strain.

To identify the dynamic stability of the single-layer ReS$_2$ under strain, we calculated the phonon spectra of the structure under different straining conditions. It is found that the single-layer ReS$_2$ is dynamically stable under a wide range of $x$-axial strain (from -9\% compressive strain to 12\% tensile strain), as can be seen from the phonon spectra in Figure \ref{fig:phonon}. The single-layer ReS$_2$ also maintains the dynamical stability under 10\% $y$-axial tensile strain, however, it becomes dynamically unstable under over -2\% $y$-axial compressive strain. The imaginary frequency around the $\bf{\Gamma}$ point in Figure \ref{fig:phonon}(e) suggests the long-wavelength instability of single-layer ReS$_2$ under -2\% $y$-axial compressive strain. Such a long-wavelength instability also appears for shear strains over 8\% as can be seen in Figures \ref{fig:phonon}(i-l). 

The phonon calculations also suggest an anisotropic nature of the single-layer ReS$_2$. Besides, the single-layer ReS$_2$ shows a long-wavelength instability over 2\% $y$-axial compressive strain and over 8\% shear strain. This finding agrees with the experimental observations. Lin \emph{et al.} found the diamond-shape Re$_4$ chain along $y$-axis can flip its direction under a small strain and consequently the structure transforms. However, the original and transformed structures correspond to the same lattice, and only differ by the choice of basis vectors \cite{lin2015single}.

\subsection{Electronic structure and photocatalytic application}

\begin{figure}[!h]
\centering
\includegraphics[width=1\textwidth]{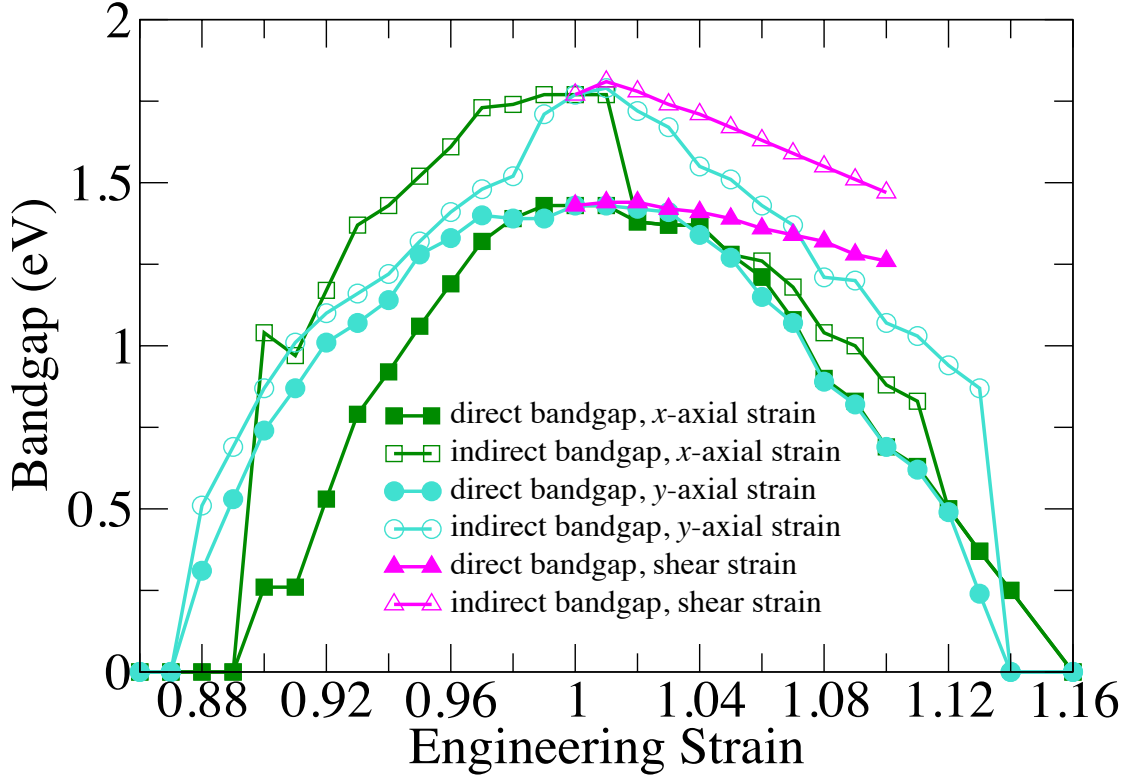}
\caption{Bandgap of single-layer ReS$_2$ under strain. \label{fig:bandgap}}
\end{figure}
\begin{figure}[!h]
\centering
\includegraphics[width=1\textwidth]{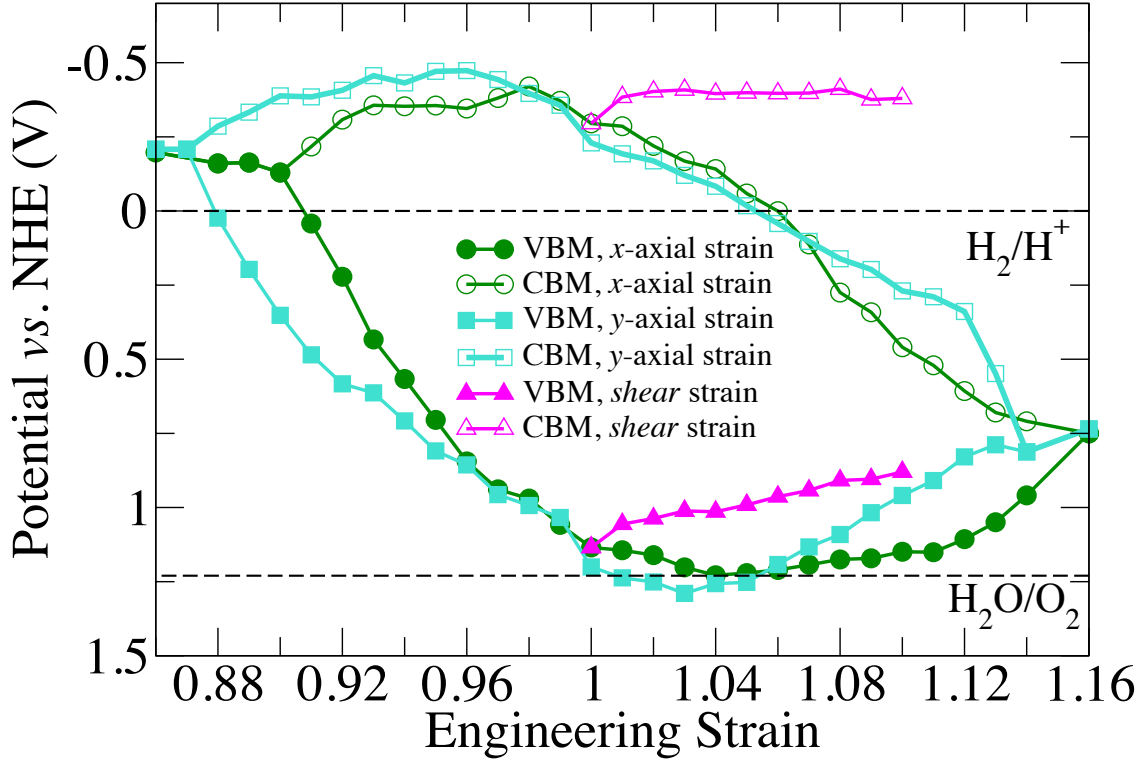}
\caption{The changes of band edge levels of single-layer ReS$_2$ under strain. \label{fig:bandedge}}
\end{figure}

Figure \ref{fig:bandgap} shows the evolution of bandgap of single-layer ReS$_2$ under engineering strain. The bandgap decreases with the increasing compressive or tensile uniaxial strain, and the single-layer ReS$_2$ can become metallic when the tensile/compressive strain is over 16\%/-10\% along the $x$-axis or over 14\%/-14\% along the $y$-axis. The bandgap also decreases with the increasing shear strain, but in a much lower speed. Worth of notice is that, the single-layer ReS$_2$ becomes a direct bandgap semiconductor under 2-5\% $x$-axial tensile strain. The indirect$-$direct bandgap transition can improve the light-absorption ratio and boost the light$-$to$-$chemical energy conversion efficiency.

Figure \ref{fig:bandedge} plots the evolution of band edge levels of single-layer ReS$_2$. The VBM of unstrained single-layer ReS$_2$ is not more positive with respect to the normal hydrogen electrode potential (NHE) than the water oxidation energy level, thus single-layer ReS$_2$ cannot catalyze the oxidation reaction. But the CBM of unstrained single-layer ReS$_2$ is more negative than the water reduction energy level, and this suggests the single-layer ReS$_2$ can catalyze the reduction reaction. The compressive uniaxial strains tend to make the VBM level more negative and do not improve the photocatalytic property for water oxidation. While, a small amount of tensile uniaxial strains can make the VBM level more positive. 1$-$5\% $y$-axial tensile strain can enable the single-layer ReS$_2$ to catalyze the water oxidation reaction. The shear strain makes both the VBM and CBM more negative with respect to the NHE potential, which enhances the reduction power but makes the water-oxidation catalyzation more impossible.

The band edge alignment with respect to the water redox potentials shows that the unstrained single-layer ReS$_2$ is not suitable for the overall water splitting in pure water. 
In practical process, the redox reaction not only depends on the band structure of photocatalysis, but also depends on the pH value of the water solution. Specifically, the electrochemical potential of the oxidation and reduction reactions can be calculated from the Nernst equation \cite{chakrapani2007charge}, respectively
\begin{eqnarray} \label{eq:redox}
\rm{E}_{O_2/H_2O}^{OX} = -4.44+(-1)(+1.229)+pH\times0.0592 eV - \frac{0.0592}{4}log_{10}(pO_2) \\
\rm{E}_{H^+/H_2}^{OX} = -4.44+(-1)(+0.401)-pOH\times0.0592 eV - \frac{0.0592}{4}log_{10}(pO_2) 
\end{eqnarray}
where $\rm{p}O_2$ designates the partial pressure of oxygen in ambient conditions, -4.44 eV corresponds to the standard hydrogen electrod potential, and \rm{pOH + pH} = 14. The water redox potentials can be adjusted to suitable levels to enable the single-layer Re$_2$S as an overall water-splitting photocatalyst. When the pH value is 3.8, both the water reduction and oxidation potentials will be shifted to positions between the CBM and VBM levels of single-layer Re$_2$S. The single-layer Re$_2$S will obtain equal amounts of reduction power and oxidation power (both are $\sim$0.1 eV), which makes the single-layer Re$_2$S capable and efficient for the overall water splitting.

As discussed above, the single-layer Re$_2$S is an indirect bandgap semiconductor and fragile against the $y$-axial compressive strain. However, strain engieering can improve its photocatalytic properties. 2$-$5\% $x$-axial tensile strain can induce the indirect$-$direct bandgap transition and improve the light absorption. 1$-$5\% $y$-axial tensile strain enables it for overall water splitting. Besides, the unstrained single-layer Re$_2$S can also acquire the overall water-splitting ability and be most efficient under acidic conditions (pH=3.8).

\section{Conclusions}
We have investigated the potential of the single-layer ReS$_2$ as a photocatalyst based on DFT calculations. The single-layer ReS$_2$ is calculated to be an indirect bandgap semiconductor and incapable of overall photocatalytic water splitting. The structural anisotropy of the single-layer ReS$_2$ is reflected by the different responses to the uniaxial strains along $x$-axis and $y$-axis. It remains stable under a wide range of $x$-axial strain, but shows a long wavelength instability over 2\% $y$-axial compressive strain. These findings agree well with the recent high-quality experiments. The indirect$-$direct bandgap transition was identified under 2$-$5\% $x$-axial tensile straining, which can improve the light absorption efficiency. It is also predicted that the single-layer ReS$_2$ can be an efficient photocatalyst for overall water splitting in acidic water solutions or under 1$-$5\% $y$-axial tensile strain.

\section*{Acknowledgement}

The authors acknowledge support from the NSFC (Grant No. 11674131), Qing Lan Project, and the Priority Academic Program Development of Jiangsu Higher Education Institutions (PAPD).

\bibliographystyle{elsarticle-harv.bst}
\bibliography{res2.bib}

\end{document}